\documentclass[10pt,twocolumn,english,aps,prl,nofootinbib]{revtex4}
\usepackage[utf8]{inputenc}
\usepackage{graphicx}
\usepackage{amsmath}
\usepackage{amssymb}
\usepackage{amsfonts}
\usepackage{lscape}
\usepackage{hyperref}
\usepackage{url}
\usepackage{epsfig}
\usepackage{color}
\usepackage{bm}
\usepackage{tabularx}
\usepackage{braket}
\newcommand{\be}{\begin{equation}}
\newcommand{\ee}{\end{equation}}
\newcommand{\ba}{\begin{eqnarray}}
\newcommand{\ea}{\end{eqnarray}}

\renewcommand{\[}{\begin{equation}}
\renewcommand{\]}{\end{equation}}

\makeatother

\begin{document}

\title{Entanglement entropy of Primordial Black Holes after inflation}

\author{Lloren\c{c} Espinosa-Portal\'es}
\email{llorenc.espinosa@uam.es}

\author{Juan Garc\'ia-Bellido}
\email{juan.garciabellido@uam.es}

\affiliation{Instituto de F\'isica Te\'orica UAM-CSIC, Universidad Auton\'oma de Madrid,
Cantoblanco, 28049 Madrid, Spain}

\date{\today}

\begin{abstract}
In this paper we study the survival of entanglement of a scalar field state created during inflation. We find that there exist UV-finite subdominant contributions to the entanglement entropy per momentum mode that scale with the number of e-folds between horizon exit and the end of inflation, and depend on the logarithm of the radius of the entangling surface, which can be taken to be the Hubble sphere. We argue that this entanglement entropy allows for the formation of entangled Primordial Black Holes (PBH). We find that the entropy arising from the entanglement between PBH is small compared with their Bekenstein entropy.
\end{abstract}
\maketitle

\section{I. Introduction}
The interplay between Gravity, QFT and Quantum Information has been a subject of increasing interest in the last decades. Following the interpretation of the horizon area of a Black Hole as its entropy \cite{Bekenstein:1973ur} and the discovery of the area law of entanglement entropy in QFT \cite{Srednicki:1993im}, there have been many studies of the entanglement entropy of the vacuum in free field theories in Minkowski \cite{Calabrese:2004eu,Casini:2009sr} or in curved backgrounds. Some of its applications include the entangled nature of the quantum states arising from particle-creation scenarios such as the Hawking and Unruh effects~\cite{Harlow:2014yka}.

Black Hole physics is certainly the most studied phenomenon within this interplay, even though many questions related to this topic remain open \cite{Harlow:2014yka,Solodukhin:2011gn}. Quantum entanglement in cosmological space-times has been however less extensively treated but can have potential interesting consequences \cite{MartinMartinez:2012sg,Martin-Martinez:2014gra}. In particular, the possibility of performing CMB experiments that may prove its quantum origin has been explored \cite{Martin:2015qta,Martin:2017zxs}. In the inflationary scenario primordial cosmological perturbations arise from quantum fluctuations that are stretched out of the Hubble scale during inflation and reenter it during the radiation- or matter-dominated eras. These perturbations are the well-known seeds for structure formation in the universe. Oftentimes their quantum origin is understated because the cosmological observables at hand do not find any distinctive quantum signature. In other words, the universe may not be classical but {\em appears} classical. This apparent contradiction is called \textit{decoherence without decoherence}\cite{Polarski:1995jg}.

If the observable universe appears classical it is because it exists in a mixed state. Hence, from a quantum-mechanical point of view, it is entangled with the non-observable universe. The issue of the quantum-to-classical transition of primordial fluctuations has been addressed with various approaches, such as quantum decoherence \cite{Kiefer:2008ku, Kiefer:2006je, Kiefer:1998qe, Polarski:1995jg} or collapse models \cite{Martin:2012pea, Das:2013qwa, Das:2014ada}. It is known that the vacuum state of de Sitter space-time is entangled in a way that goes beyond the area law found in Minkowski space-time, as it was found by Maldacena and Pimentel \cite{Maldacena:2012xp}. Its corresponding entanglement entropy includes both UV-divergent and UV-finite terms. The former arise from local physics, while the latter are related to true long-range or non-local correlations. If this entanglement arises in deSitter space-time, it must be at least partially created during inflation as well. Entanglement may occur between different momentum modes as well between localized modes and it may change during time evolution, since it may not be unitary when restricted to individual modes due to interactions among them. However, the whole quantum state of the field must remain pure as dictated by unitary evolution.

We argue that this kind of entanglement may survive after inflation. In particular, we explore in this paper how some terms can be related to the entanglement of isotropic modes across a spherical entangling surface. The radius of this sphere is arbitrary in our formalism, but we will discuss in more detail the case of the Hubble sphere, given its physical interest. This entanglement would also affect a Primordial Black Hole (PBH) formed by the gravitational collapse of a casual domain during the radiation era \cite{GarciaBellido:1996qt}. We would like to understand how this entanglement entropy among the PBH and the rest of the universe may act as an {\em entanglement trap} among the PBH themselves and preserve a long-range correlation between them. In a future work we will explore the phenomenological consequences of this correlation.

The paper is organized as follows. In sections II and III we set the context for our work by briefly reviewing the concept of quantum entanglement in field theories and the time evolution of the vacuum state during inflation in the Schrödinger picture. In section IV we reformulate the theory of a scalar field in a radiation-dominated universe by means of canonical quantization in spherical coordinates. In section V we compute the entanglement entropy per mode of the resulting state. In section VI we discuss how the bipartition of a quantum mode in an inner and an outer component works. In section VII we show how the contributions for each mode should be integrated to give the final result. In section VIII we make some comments about the implication of these results for Primordial Black Hole formation. Finally we sum up with conclusions and outlook to further research.

\section{II. Entanglement in field theory}

Entanglement is the phenomenon by which correlations of quantum origin can arise between observations of different physical systems. Such systems are actually subsystems of a larger quantum system and they cannot be described separately by a vector of its corresponding Hilbert space, what is called a pure state, but rather as a statistical ensemble of possibly many such vectors, what is called a mixed state. Should the larger quantum system not be entangled with other systems, then it can be described by a vector on the tensor product of the Hilbert spaces of each of the subsystems. The perhaps paradigmatic example of an entangled quantum system is the singlet state of a system of two particles with spin-$1/2$:

\begin{equation}
\ket{\psi} = \frac{1}{\sqrt{2}} \left( \ket{+-} - \ket{-+} \right)
\end{equation}

The density matrix of this state is given by $\rho = \ket{\psi}\bra{\psi}$. Then the reduced density matrix that describes the mixed state of each subsystem is obtained by tracing over the other subsystem:

\begin{equation}
\rho_i = \textrm{Tr}_j \rho
\end{equation}

And the entanglement entropy is given by the von Neumann entropy of the mixed state of any of the subsystems:

\begin{equation}
S_{\textrm{ent}} = - \textrm{Tr} \left(\rho_i \log \rho_i \right)
\end{equation}

Quantum entanglement also occurs when dealing with continuously infinite degrees of freedom. In fact, it is an inevitable and natural feature of any quantum field theory. If we take the whole field to be the quantum system of interest, then it can be split into subsystems whose correlations are measured by their entanglement entropy. This entanglement entropy is dependent on the quantum state of the field and the choice of subsystems. For instance, if we consider the vacuum state of a scalar field theory in Minkowski space-time, it can be expressed as a product state of single momentum mode vacua and therefore there is no entanglement between them:

\begin{equation}
\ket{0} = \otimes_k \ket{0}_k
\end{equation}

However, if we choose the subsystems to be the localized modes inside and outside of a sphere of radius $R$, then one finds quantum entanglement between the inner and the outer modes with a UV-divergent entanglement entropy that scales with the area of the sphere \cite{Srednicki:1993im}:

\begin{equation}
S \sim \Lambda^2 R^2
\end{equation}

Where $\Lambda$ is the UV energy cut-off. This is the celebrated area law and describes the dominant contribution to the entanglement entropy of the vacuum state in Minkowski space-time. It is interpreted as the entanglement of particles close to the surface of the sphere and is therefore related to local physics \cite{Maldacena:2012xp}.

Other quantum states may exhibit for instance quantum entanglement between different momentum modes. Non-trivial gravitational backgrounds have also an effect on the entanglement entropy and may add both UV-divergent and UV-finite contributions beyond the area law. For a massless free minimally coupled scalar field in de Sitter space-time these are given by \cite{Maldacena:2012xp}:

\begin{equation}
\begin{aligned}
& S_{\textrm{dS, UV-divergent}} =  c_1 \Lambda^2 A + \log \left(\Lambda^{-1} H \right) \left( c_2 + c_3 A H^2 \right)\\
& S_{\textrm{dS, UV-finite}} = c_4 A H^2 + c_5 \log\left(-\eta\right) + \textrm{constant}
\end{aligned}
\end{equation}

The term $\sim \log\left(-\eta\right)$ signals the presence of long-range quantum correlations. They arise from short-range physics due to the streching out of length-scales with the expansion. Since during inflation the background metric can be regarded as approximate de Sitter space-time, we argue that such long-range quantum correlations may also be created during inflation and survive during the subsequent radiation-dominated era.

\section{III. The quantum state after inflation}

We consider a massless field $\Phi$ that can be used for instance to describe primordial curvature perturbations. Since primordial gravitational waves are described by the same dynamics, our results will also be valid for them. The choice of the vacuum state in dS is not unique due to the lack of a time-like Killing vector. A possible criterion to fix the vacuum state is to pick the mode functions so that they reduce to plane waves in the distant past. This defines the so called Bunch-Davies vacuum, which is usually accepted to be the most reasonable option \cite{Mukhanov:2007zz}, even though alternatives exist and have been studied \cite{Allen:1985ux, Danielsson:2002qh, Brunetti:2005pr}. It is specially safe to choose the Bunch-Davies vacuum in applications to inflation, since only a piece of dS is actually needed to describe a short period of accelerated expansion and those modes with wavelength larger than the event horizon at the beginning of inflation are phenomenologically irrelevant.

In the Schr\"odinger picture, the Bunch-Davies vacuum evolves into a squeezed state due to the action of the time-evolution operator (S-matrix):

\begin{equation}
S (\eta) = e^{-i\eta H(\eta)}
\end{equation}

With a Hamiltonian that contains a squeezing term:

\begin{equation}
\begin{aligned}
H = & \frac{1}{2} \int d^3k  \bigg[ k \left(a(\vec{k},\eta) a^{\dagger}(\vec{k},\eta) \right) + \\
& + i \frac{a'}{a} \left( a^{\dagger}(\vec{k},\eta)a^{\dagger}(\vec{-k},\eta) - a(\vec{k},\eta)a(\vec{-k},\eta) \right) \bigg]
\end{aligned}
\end{equation}

It can be shown that the time-evolution operator can be rewritten in the following way \cite{Albrecht:1992kf}:

\begin{equation}
\begin{aligned}
\log S(\eta) =  & \int d^3k \frac{\tau(\vec{k},\eta)}{2}\bigg[a(\vec{k},\eta_0)a(\vec{-k},\eta_0)e^{-i\phi(\vec{k},\eta)}\\
& -a^{\dagger}(\vec{k},\eta_0)a^{\dagger}(\vec{-k},\eta_0)e^{i\phi(\vec{k},\eta)} \bigg]
\end{aligned}
\end{equation}

And it acts on the vacuum creating a two-mode squeezed state, which entangles the $\vec{k}$ and $-\vec{k}$ modes:

\begin{equation}
\begin{aligned}
\ket{0,\eta} & = S(\eta) \ket{0,\eta_0}\\
& = \otimes_k \frac{1}{\cosh \tau} \sum_{n=0}^{\infty}\left(e^{-i\phi} \tanh \tau \right)^n \ket{n}_{\vec{k}}\ket{n}_{-\vec{k}}
\end{aligned}
\end{equation}

Where $\eta_0$ is the conformal time at the beginning of inflation and $\tau$ and $\phi$ are respectively the squeezing parameter and phase, which depend only on the conformal time $\eta$ and the norm of the momentum $k$. We refer the reader to \cite{Schumaker} for a review of the physics and mathematics of squeezed states as well as to the original references on two-mode squeezed states \cite{Caves:1985zz,Schumaker:1985zz}. In the problem at hand one finds that $\tau \sim N$ where $N$ is the number of e-folds between horizon exit and the end of inflation.

This state shows entanglement between $\vec{k}$ and $-\vec{k}$ modes and its  entanglement entropy is given by \cite{Prokopec:1992ia}:

\begin{equation}
S_{ent} = 2\left[\log(\cosh \tau) - \log (\tanh \tau) \sinh^2 \tau \right]
\end{equation}

Which reduces to $S_{ent} \simeq 2\tau$ for $\tau \gg 1$ as is usually the case. This entanglement entropy is related to the coarsed-grained entropy of primordial perturbations computed by Brandenberger, Mukhanov and Prokopec \cite{Brandenberger:1992jh}. Indeed they found the entropy density to be:

\begin{equation}
s = \int d^3k \log \sinh^2 \tau_k \simeq \int d^3k 2 \tau_k
\end{equation}

It is true that apparently we are comparing entropy density with total entropy, but it is not the case since after integrating the entanglement entropy over all possible momentum modes we get a quantity in units of entropy density. The scaling can be properly regularized via discretization:

\begin{equation}
\int d^3k \rightarrow \sum_k = \left(\frac{k_{max}}{k_{min}} \right) \sim k_{max}^3 L^3
\end{equation}

Which indeed grows as the volume.

It would be interesting to check other ways in which quantum entanglement is present in this state. In particular we will try to ellucidate the entanglement between modes restricted to the interior and the exterior of a sphere of radius $R$.

\section{IV. Canonical quantization in spherical coordinates} 
Introducing the auxiliary field $\chi = a \Phi$ the equation of motion of the scalar field takes a simple form \cite{Mukhanov:2007zz}:

\begin{equation}
\chi '' - \nabla^2 \chi - \frac{a''}{a} \chi = 0
\end{equation}

Using the fact that during the radiation-dominated era $a \sim \eta$ the equation of motion reduces to that on Minkowski space-time and therefore its solutions are the well-known plane waves. In spherical coordinates this is equivalent to:
\begin{equation}
	\frac{\partial^2\chi}{\partial\eta^2} - \frac{1}{r}\frac{\partial^2}{\partial r^2}\left(r \chi\right) - \frac{1}{r^2}\Delta_{S_2}\chi = 0
\end{equation}

Where the Laplacian on the 2-sphere is given by:

\begin{equation}
\Delta_{S_2} = \frac{1}{\sin \theta} \frac{\partial}{\partial \theta} \left( \sin \theta \frac{\partial}{\partial \theta} \right) + \frac{1}{\sin^2 \theta} \frac{\partial^2}{\partial \varphi^2}
\end{equation} 

The solutions to this equation are known to be:

\begin{equation}
	\chi_{k,l,m}(\eta, r,\theta,\varphi) = \frac{1}{\sqrt{2\omega}}e^{-i\omega \eta} j_l(kr)Y_{lm}(\theta,\varphi)
\end{equation}

Where $j_l(z) = \sqrt{\frac{\pi}{2z}}J_{l+1/2}(z)$ are the spherical Bessel functions and $Y_{lm}(\theta,\varphi)$ are the spherical harmonics. Notice that for a massless field, as it is our case, the dispersion relation is $\omega = k$.

We need to normalize this with respect to the Klein-Gordon inner product.

\begin{equation}
\begin{aligned}
	\left( \chi_{klm}, \chi_{k'l'm'}\right) & = i \int_0^{\infty}r^2dr \int d\Omega \left( \chi^*_{klm} \overset{\leftrightarrow}{\partial_{\eta}} \chi_{k'l'm'} \right)\\& = \frac{\pi}{2k^2}\delta(k-k')\delta_{ll'}\delta_{mm'}
\end{aligned}
\end{equation}

The choice of functions makes therefore perfect sense from the point of view of the Klein-Gordon inner product, since they are orthogonal. We reabsorb the factor $1/k^2$ into the definition of the mode functions since we anticipate it to be important for the operator field expansion. We also reabsorb the constant factor $\pi/2$.

\begin{equation}
\chi_{klm}(\eta,r,\theta,\varphi) = \frac{1}{\sqrt{2\omega}}e^{-i\omega\eta} \sqrt{\frac{2}{\pi}} k j_l(kr) Y_{lm}(\theta,\varphi)
\end{equation}

The field operator $\chi$ can be expanded in terms of these functions:

\begin{equation}
\begin{aligned}
\chi(\eta,r,\theta,\varphi) = & \int_0^{\infty}dk \sum_{l=0}^{\infty} \sum_{m=-l}^l \frac{k}{\sqrt{2\omega}} j_l(kr) \cdot \\
& \cdot \left(Y^*_{lm}(\theta,\varphi)e^{i\omega\eta} a_{klm}+Y_{lm}(\theta,\varphi)e^{-i\omega\eta}a^{\dagger}_{klm}\right)
\end{aligned}
\end{equation}

The field operator must of course satisfy the Canonical Commutation Relation:

\begin{equation}
[\chi(\eta,r,\theta,\varphi),\Pi(\eta,r',\theta',\varphi')] = i\delta^{(3)}(\vec{r}-\vec{r}')
\end{equation}

Which is achieved by imposing:

\begin{equation}
\begin{aligned}
\left[a_{klm},a_{k'l'm'}\right] = 0 = \left[a^{\dagger}_{klm},a^{\dagger}_{k'l'm'}\right] \\
\left[a_{klm},a^{\dagger}_{k'l'm'}\right] = \delta(k-k')\delta_{ll'}\delta{mm'}
\end{aligned}
\end{equation}

As one would expect, this canonical quantization in spherical coordinates is completely equivalent to the usual canonical quantization in cartesian coordinates. The destruction and creation operators in both descriptions are related by the following expression:

\begin{equation}
a_{\vec{k}} = \sum_{l=0}^{\infty} \sum_{m=-l}^l \frac{i^l}{k} Y_{lm}(\hat{k}) a_{klm}
\end{equation}

And its inverse:

\begin{equation}
a_{klm} = (-i)^l k \int d\Omega Y^*_{lm}(\hat{k}) a_{\vec{k}}
\end{equation}

Where $\hat{k} = \vec{k}/k$ and is simply parametrized by two angular variables. In terms of this creation and annihilation operators in spherical coordinates the time-evolution operator becomes:

\begin{equation}
\begin{aligned}
\log S(\eta) = & \int d^3k \frac{\tau(\vec{k},\eta)}{2} \sum_{l,l',m,m'}\\
& \bigg[\frac{i^{l+l'}}{k^2} Y_{lm}(\hat{k})Y_{l'm'}(-\hat{k})a_{klm}a_{kl'm'} e^{-i\phi(\vec{k},\eta)}\\
& - \frac{(-i)^{l+l'}}{k^2} Y^*_{lm}(\hat{k})Y^*_{l'm'}(-\hat{k})a^{\dagger}_{klm}a^{\dagger}_{kl'm'}e^{i\phi(\vec{k},\eta)} \bigg]
\end{aligned}
\end{equation}

After applying some properties of the spherical harmonics and integrating over the angular variables one gets a simpler expression for the operator:

\begin{equation}
\begin{aligned}
\log S(\eta) = & \int dk \frac{\tau}{2} \sum_{l,m}\\
& (-1)^{m} \cdot \bigg[a_{klm}a_{kl,-m} e^{-i\phi} - a^{\dagger}_{klm}a^{\dagger}_{kl,-m}e^{i\phi} \bigg]
\end{aligned}
\end{equation}

This operator has a slightly different effect for $l=0$ and $l\neq 0$. Indeed by expressing:

\begin{equation}
S(\eta) = \prod_{l,m} S_{lm}(\eta)
\end{equation}

We see that:

\begin{equation}
\begin{aligned}
& \log S_{00} (\eta) =\\
& \int dk \frac{\tau(\vec{k},\eta)}{2} \left[ a_{k00} a_{k00} e^{-i\phi(\vec{k},\eta)} - a^{\dagger}_{k00} a^{\dagger}_{k00} e^{i\phi(\vec{k},\eta)} \right]
\end{aligned}
\end{equation}

The operator $S_{00}$ creates nothing but a one-mode squeezed operator out of the vacuum. By factoring the state as well:

\begin{equation}
	\ket{0,\eta} = \otimes_{lm} \ket{0,\eta}_{lm}
\end{equation}

We find that:

\begin{equation}
\begin{aligned}
	\ket{0,\eta}_{00} & = S_{00}(\eta) \ket{0}\\
	& = \otimes_{k^2} \frac{1}{\sqrt{\cosh \tau}} \sum_{n=0}^{\infty} \frac{\sqrt{(2n!)}}{n!}\left( -\frac{1}{2}e^{2i\phi} \tanh \tau \right) \ket{2n}_{k00}
\end{aligned}
\end{equation}

On the other hand, for the other modes $S_{lm}$ is a two-mode squeezing operator:

\begin{equation}
\begin{aligned}
\log S_{lm} (\eta) = & \int dk \frac{\tau}{2} (-1)^m\\
&\left[ a_{klm} a_{kl,-m} e^{-i\phi(\vec{k},\eta)} - a^{\dagger}_{klm} a^{\dagger}_{kl,-m} e^{i\phi(\vec{k},\eta)} \right]
\end{aligned}
\end{equation}

Which creates a two-mode squeezed state. This kind of state carries entanglement between the $m$ and $-m$ modes:

\begin{equation}
	\ket{0,\eta}_{lm} = \otimes_{k^2} \sum_{n=0}^{\infty} \frac{\left(e^{2i\phi}(-1)^{m+1}\tanh \tau \right)^n}{\cosh \tau} \ket{n}_{klm} \ket{n}_{kl,-m} 
\end{equation}

To sum up, in spherical coordinates the quantum state after inflation has the following properties:

\begin{itemize}
	\item The isotropic mode $l=0$ is found in a one-mode squeezed state.
	\item The anisotropic modes $l\neq0$ are found  in a two-mode squeezed state, which entangles $m$ and $-m$ modes. This is one source of entanglement, but there is still another one due to the in and out bipartition by a spherical entangling surface of radius $R$.
\end{itemize}

\section{V. Computing the entanglement entropy}

As stated in the previous section, the anisotropic modes (i.e. those with $l\neq 0$) are found in two-mode squeezed states and show therefore entanglement between $m$ and $-m$ modes. This entanglement is related directly to the entanglement between $\vec{k}$ and $-\vec{k}$ modes that is found in cartesian coordinates. The computation of its entanglement entropy follows analagously and delivers the same result $S_{ent} \simeq 2\tau$ for large $\tau$.

The second simplest form of entanglement is the one across a spherical entangling surface of radius $R$ for isotropic modes, i.e. those with $l=0$. This entanglement is most interesting when $R$ is taken to be the Hubble radius, but we will keep it as a free parameter for now. We will proceed with the ansatz that the creation and destruction operators can be split into an inner and an outer component as follows:

\begin{equation}
a_{k00} \equiv a_k = \alpha a_{k,in} + \beta a_{k,out}
\end{equation}

With $|\alpha|^2 + |\beta|^2 = 1$ and the usual Canonical Commutation Relations (CCR), in which it should be taken into account that the inner and outer operators commute:

\begin{equation}
\begin{aligned}
&\left[a_{k,in},a_{k',in}\right] = \left[a_{k,out},a_{k',out}\right] = 0\\
&\left[a^{\dagger}_{k,in},a^{\dagger}_{k',in}\right] = \left[a^{\dagger}_{k,out},a^{\dagger}_{k',out}\right] = 0\\
&\left[a_{k,in},a_{k',out} \right] = \left[a^{\dagger}_{k,in},a^{\dagger}_{k',out} \right] = \left[a_{k,in},a^{\dagger}_{k',out}\right] = 0
\end{aligned}
\end{equation}

We will deal later with the fact that, in general, the following commutators do not satisfy the canonical relations:

\begin{equation}
\left[a_{k,in},a^{\dagger}_{k',in}\right] \neq \delta (k-k') \neq \left[a_{k,out},a^{\dagger}_{k',out}\right]\\
\end{equation}

Any quantum state can be expressed in terms of n-particle states created by these inner and outer operators, which take the following form:

\begin{equation}
\begin{aligned}
\ket{n} & = \frac{1}{\sqrt{n!}} \left(a^{\dagger}\right)^n \ket{0} = \left(\alpha^* a^{\dagger}_{in} + \beta^* a^{\dagger}_{out}\right)^n \ket{0}\\
& = \sum_{m=0}^n \binom{n}{m}^{1/2} \alpha^m \beta^{n-m} \ket{m}_{in} \otimes \ket{n-m}_{out}
\end{aligned}
\end{equation}

Now, the $l=0$ sector of the vacuum state is a one-mode squeezed state, which can be written in its standard particle basis decomposition and then split into inner and outer components:

\begin{equation}
\begin{aligned}
\ket{0,\eta}_{00} &  = \frac{1}{\sqrt{\cosh \tau}} \sum_{n=0}^{\infty} \frac{\sqrt{(2n)!}}{n!} \left( -\frac{1}{2} e^{2i\phi} \tanh \tau \right)^n \cdot\\
&\cdot \sum_{m=0}^{2n} \binom{2n}{m}^{1/2} \alpha^m \beta^{n-m}\ket{m}_{in}\otimes \ket{2n-m}_{out}
\end{aligned}
\end{equation}

And we can build the corresponding density matrix:

\begin{equation}
\begin{aligned}
\rho_{00} & =  \ket{0,\eta}_{00} \bra{0,\eta}_{00}\\
& = \frac{1}{\cosh \tau} \sum_{n,n'=0}^{\infty} (-2)^{-(n+n')} \frac{\sqrt{(2n)!(2n')!}}{n! n'!} \cdot\\
& \cdot e^{2i\phi(n-n')} \tanh^{n+n'} (\tau)  \cdot \sum_{m,m'=0}^{2n,2n'} \binom{2n}{m}^{1/2} \binom{2n'}{m'}^{1/2} \cdot\\
& \cdot \alpha^{m+m'} \beta^{(n-m)+(n'-m')}\\
& \cdot \ket{m}_{in} \bra{m'}_{in} \otimes  \ket{2n-m}_{out} \bra{2n'-m'}_{out}
\end{aligned}
\end{equation}

Now we trace out the inner degrees of freedom in order to obtain the reduced density matrix of the outer degrees of freedom.

\begin{equation}
\begin{aligned}
\rho_{out} & = \textrm{Tr}_{in} \rho = \sum_{q=0}^{\infty} \bra{q}_{in} \rho \ket{q}_{in}\\
& = \frac{1}{\cosh \tau}\sum_{n,n'=0}^{\infty} \sum_{l=0}^{\min(2n,2n')}(-2)^{-(n+n')} \frac{\sqrt{(2n)!(2n')!}}{n! n'!}\\
& e^{2i\phi(n-n')} \tanh^{n+n'} (\tau) \alpha^{2l} \beta^{n+n'-2l}  \cdot\\
& \cdot \binom{2n}{l}^{1/2}\binom{2n'}{l}^{1/2} \ket{2n-l}_{out}\bra{2n'-l}_{out}
\end{aligned}
\end{equation}

In order to compute the von Neumann entropy of this density matrix we would in principle need to compute its logarithm and, therefore, diagonalize it. Its complicated structure and infinite size make it seem an impossible task. Hence, we will compute it using a different method, namely exploiting the available knowledge of the von Neumann entropy of generic two-mode Gaussian states. Even though it may not seem obvious that $\rho_{00}$ is a Gaussian state, it has been proven that any quantum state created by a time evolution driven by a bilinear two-mode Hamiltonian is a two-mode Gaussian state \cite{Schumaker}. This means that, even though the state itself is characterized by an infinite set of coefficients, it only contains a much more reduced amount of information codified in its first and second statistical moments, that is, in its expected values and covariance matrix. In other words: the density matrix of a single mode is created from the vacuum by acting with a squeezing operator, which depends on a few parameters, two per momentum mode. Therefore, its entanglement entropy should also depend on these parameters only. This means that, even though one needs in principle all the matrix elements to compute the logarithm of the matrix, it cannot have any non-trivial dependence that is not encoded in the dependence on the parameters. We use in the following the formalism described in \cite{Serafini:2003ke} to compute the entanglement entropy.

We introduce the following auxiliary field and conjugated momentum operators:

\begin{equation}
\begin{aligned}
&\chi_{in/out} = \frac{1}{\sqrt{2}} \left(a_{in/out} + a^{\dagger}_{in/out} \right)\\
&\pi_{in/out} = \frac{-i}{\sqrt{2}} \left(a_{in/out} - a^{\dagger}_{in/out} \right)
\end{aligned}
\end{equation}

Then one defines the covariance matrix $\sigma$ of a quantum state as follows:

\begin{equation}
\sigma_{ij} = \frac{1}{2} \left< x_i x_j + x_j x_i \right> - \left<x_i \right> \left<x_j \right> 
\end{equation}

Where $i=1,2$ and the vector $x$ is defined as $x = ( \chi_{in}, \pi_{in})^T $

The expected values $\left<x_i \right>$ can be set to zero without loss of generality. As a matter of fact, they are zero in our case. Let us use the short notation:

\begin{equation}
\rho_{out} = \sum_{n,n'=0}^{\infty} \sum_{l=0}^{\min(2n,2n')} c_{nn'l} \ket{2n-l}_{out} \bra{2n'-l}_{out}
\end{equation}

Then:

\begin{equation}
\begin{aligned}
\left< a^{\dagger}_{out} \right> & = \textrm{Tr}\left(\rho_{out}a^{\dagger}_{out} \right)\\
& = \sum_{n,n'=0}^{\infty} \sum_{l=0}^{max(2n,2n')} c_{nn'l} \sqrt{2n'-l} \cdot \delta_{2n-l,2n'-l-1}\\
& = 0
\end{aligned}
\end{equation}

This is $0$ because the condition of the Kronecker delta can never be fulfilled since $n$ and $n'$ are integers. Similarly one obtains $\left< a_{out} \right> = 0$. Hence, we focus on the second statistical moments:

\begin{widetext}

\begin{equation}
\begin{aligned}
	\left< a^{\dagger} a \right> & = \textrm{Tr} \left( \rho a^{\dagger} a\right) = \textrm{Tr} \left( a \rho a^{\dagger} \right)\\ & = \textrm{Tr}\left( \sum_{n,n'=0}^{\infty} \sum_{l=0}^{\min(2n,2n')} c_{nn'l} \sqrt{(2n-l)(2n'-l)} \ket{2n-l-1}\bra{2n'-l-1} \right)\\
	& =\sum_{n,n'=0}^{\infty} \sum_{l=0}^{\min(2n-1,2n'-1)} c_{nn'l} \sqrt{(2n-l)(2n'-l)} \delta_{2n-l-1,2n'-l-1} \\
	& = \frac{1}{\cosh \tau} \sum_{n=0}^{\infty} (-2)^{-2n} \frac{(2n)!}{(n!)^2} \tanh^{2n} \tau \cdot 2n \sum_{l=0}^{2n-1}\frac{(2n-1)!}{l!(2n-l-1)!} \alpha^{2l} \beta^{2(n-l)}\\
	& = \frac{1}{\cosh \tau} \sum_{n=0}^{\infty} 2^{-2n} \frac{(2n)!}{(n!)^2} \tanh^{2n} \tau \cdot 2n \beta^2 = \beta^2 \sinh^2 \tau\\
\end{aligned}
\end{equation}

And the same for the other moment:

\begin{equation}
\begin{aligned}
	\left<aa\right> & = \textrm{Tr} \left( \rho a a\right) = \textrm{Tr} \left( a \rho a \right)\\
	& = \textrm{Tr} \left( \sum_{n,n'=0}^{\infty} \sum_{l=0}^{\min(2n-1,2n')} c_{nn'l} \cdot \sqrt{(2n'-l+1)(2n-l)} \cdot \ket{2n-l-1}\bra{2n'-l+1} \right)\\
	& = \sum_{n,n'=0}^{\infty} \sum_{l=0}^{\min(2n-1,2n')} c_{nn'l} \cdot \sqrt{(2n'-l+1)(2n-l)} \cdot \delta_{2n-l-1,2n'-l+1}\\
	& = \frac{1}{\cosh \tau} \sum_{n=1}^{\infty} \sum_{l=0}^{2n-2} \sqrt{(2n-l-1)(2n-l)} (-2)^{-2n+1} \frac{\sqrt{(2n)!(2n-2)!}}{n!(n-1)!} \cdot\\
	& \cdot e^{2i\phi} \tanh^{2n-1} \tau \binom{2n}{l}^{1/2} \binom{2n-2}{l}^{1/2} \alpha^{2l}\beta^{4n-2l-2}\\
	& = \frac{1}{\cosh \tau} \sum_{n=1}^{\infty}  2^{-2n+1}\frac{(2n)!}{n!(n-1)!} e^{2i\phi} \tanh^{2n-1} \tau\sum_{l=0}^{2n-2} \binom{2n-2}{l} \alpha^{2l}\beta^{4n-2l-2}\\
	& = \frac{1}{\cosh \tau} \sum_{n=1}^{\infty} 2^{-2n+1} \frac{(2n)!}{n!(n-1)!} e^{2i\phi} \beta^2 \tanh^{2n-1} \tau = e^{2i\phi} \beta^2\sinh \tau\cosh \tau
\end{aligned}
\end{equation}

\end{widetext}

We will neglect in the following the contribution of the phase, since we can always reabsorb it by means of the transformation $a \rightarrow e^{-i\phi}a$ which does not affect the physics of the problem.

With this we can compute the elements of the covariance matrix:

\begin{equation}
\begin{aligned}
\sigma_{\chi \chi} & = \left< \chi \chi \right> = \beta^2 e^\tau \sinh \tau + \frac{1}{2}
\end{aligned} 
\end{equation}

\begin{equation}
\begin{aligned}
\sigma_{\pi \pi} & = \left< \pi \pi \right> = \frac{1}{2} - \beta^2 e^{-\tau} \sinh \tau
\end{aligned}
\end{equation}

\begin{equation}
\begin{aligned}
\sigma_{\chi \pi} & = 0
\end{aligned}
\end{equation}

The entanglement entropy of the quantum state is related to the determinant of the covariance matrix as follows:

\begin{equation}
S = \frac{1-\mu}{2\mu} \ln \left(\frac{1+\mu}{1-\mu}\right) - \ln \left( \frac{2\mu}{1+\mu} \right)
\end{equation}

With:

\begin{equation}
\mu = \frac{1}{2^n \sqrt{\det \sigma}}
\end{equation}

Where $n$ is the number of quantum modes. In our present case, $n=1$. And the determinant is given by:

\begin{equation}
\begin{aligned}
\det \sigma & = \sigma_{\chi \chi} \sigma_{\pi \pi} - \sigma_{\pi\chi}^2 = \frac{1}{4} + \beta^2 (1-\beta^2) \sinh^2 \tau
\end{aligned}
\end{equation}

Notice that this result is symmetric under the exchange of $\beta^2$ and $\alpha^2 = 1- \beta^2$ This consistency requirement is of uttermost importance.

And so:

\begin{equation}
\mu = \frac{1}{2 \sqrt{\det \sigma}} = \frac{1}{\sqrt{1+4\beta^2 \alpha^2 \sinh^2 \tau}}
\end{equation}

Being the result for the entanglement entropy:

\begin{equation}
\begin{aligned}
S = & \log \left[\frac{1}{2} \left(1+\sqrt{1+4\beta^2 \alpha^2 \sinh^2 \tau}\right)\right]\\
&+\frac{1}{2}\left(-1+\sqrt{1+4\beta^2\alpha^2 \sinh^2 \tau}\right)\cdot \\
& \cdot \log \left( \frac{1+\sqrt{1+4\beta^2 \alpha^2 \sinh^2 \tau}}{-1+\sqrt{1+4\beta^2 \alpha^2 \sinh^2 \tau}}\right)
\end{aligned}
\end{equation}

Now, let us consider a completely equal bipartiton, i.e. $\alpha = \beta = \frac{1}{\sqrt{2}}$. Then:

\begin{equation}
\mu = \frac{1}{\sqrt{1+\sinh^2 \tau}} = \textrm{sech} \tau
\end{equation}

For $\tau \gg 1$, as it is usually the case in cosmological applications, this in turn leads to:

\begin{equation}
S \sim 2\tau
\end{equation}

Which means that the entanglement between inner and outer modes grows linearly with $\tau$ and vanishes for $\tau=0$. This turns out to be the case as well for any other value of $\beta$. The main difference is that the linear behaviour is preceded by a slow exponential growth before becoming linear, and the more $\beta$ departs from its equipartion value $\beta = 1/\sqrt{2}$ the longer this linear behavior appears.

On the other hand, for fixed $\tau$ the following dependence for $\alpha < \frac{1}{\sqrt{2}}$ is observed:

\begin{equation}
S \sim \log \alpha
\end{equation}

We will discuss this in a later section but we advance the following Ansatz for the scaling of the coefficients $\alpha$ and $\beta$:

\begin{equation}
\alpha = \sqrt{\frac{R}{L}} \quad \textrm{and} \quad \beta = \sqrt{1-\frac{R}{L}}
\end{equation}

So that:

\begin{equation}
S \sim \log\frac{R}{L}
\end{equation}

Where $L$ is an IR regulator. Therefore, an IR divergence arises due to the term $\log L$. But actually for \textit{really small} $\alpha$ we have that $S\rightarrow 0$. This can be checked taking the complete formula or, more easily, performing a Taylor expansion around $\alpha = 0$:

\begin{equation}
S \simeq \alpha^2 \left[1-\log \left(\alpha^2\sinh^2 \tau\right)\right]\sinh^2 \tau
\end{equation}

This result should be interpreted carefully. Indeed, if we take the limit $L\rightarrow \infty$ this is in a sense equivalent to taking the limit $R \rightarrow 0$. This would mean that all degrees of freedom have been traced out and so the entanglement entropy must vanish. The actual quantity should be regularized. We think a reasonable regularization scheme would be taking the Hubble scale during inflation as initial size of the universe and then expand it exponentially during the $N$ e-folds that inflation lasts:

\begin{equation}
L = H^{-1} e^N
\end{equation}

This prescription is borrowed from regularization schemes in quantum cosmology and stochastic inflation \cite{GarciaBellido:1993wn,GarciaBellido:1994ci,Garriga:2001ri}. It is also consistent with the Bunch-Davies prescription for the vacuum state, since it cannot be applied to modes whose wavelength was larger than the Hubble scale at the beginning of inflation.

The key is that in any case it scales as $S \sim \log R$. This does not however violate the area law, because this form of entanglement arises solely due to the squeezing and vanishes the moment the limit $\tau \rightarrow 0$ is taken. The usual short-range UV-divergent and area-scaling contribution to the entanglement entropy must still be present when the total entanglement entropy is computed but is not related to the isotropic modes. From our expression it can be inferred that the entanglement entropy given by the long-range correlations between isotropic modes is in any case subdominant. However, to make a proper judgement it should still be integrated for all the available modes.

\section{VI. Mode bipartition}

The expression we used to split the creation and annihilation operators of the scalar field theory defined on the whole space-time manifold seems a bit obscure. In this section we will argue why the coefficients $\alpha$ and $\beta$ should scale as indicated before.

In order to do this, let us place the theory in a spherically symmetric lattice, so that the radial coordinate is discretized while keeping the angular coordinates continuous. Then the field itself is discretized into a set of fields $\chi_r(\theta,\varphi)$ living at each point of the lattice and can be expanded in terms of its associated annihilation and creation operators $a_r$ and $a^{\dagger}_r$. They satisfy the canonical commutation relations:

\begin{equation}
\left[ a_r, a^{\dagger}_{r'} \right] \sim \delta_{rr'}
\end{equation}

Or, in the continuum limit:

\begin{equation}
\left[ a_r, a^{\dagger}_{r'} \right] = \frac{1}{4\pi r^2} \delta(r-r')
\end{equation}

The usual momentum-defined creation and annihilation operators are recovered through a Bessel transform in the continuum limit:

\begin{equation}
a_k = \int d^3r \sqrt{\frac{2}{\pi}} j_0(kr) a_r
\end{equation}

We can split this integral into two regions and so define the inner and outer components of the operator:

\begin{equation}
a_k = 4\pi \int_0^R dr r^2 k \sqrt{\frac{2}{\pi}} j_0(kr) a_r + 4\pi \int_R^{\infty} dr r^2 k \sqrt{\frac{2}{\pi}} j_0(kr) a_r
\end{equation}

And we can approximately identify:

\begin{equation}
\begin{aligned}
a_{k,in} \sim 4\pi \int_0^R dr r^2 k \sqrt{\frac{2}{\pi}} j_0(kr) a_r\\
a_{k,out} \sim 4\pi \int_R^{\infty} dr r^2 k \sqrt{\frac{2}{\pi}} j_0(kr) a_r
\end{aligned}
\end{equation}

The integrals are defined in three dimensions and the delta functions is defined to be the spherically symmetric three dimensional one in order to show that this formalism can be generalized to include anistropic modes, even though we will not need them here.

From this point of view it is clear that it is legitimate to perform a bipartition of the local degrees of freedom of the scalar field into inner and outer components with respect to some spherical surface of radius $R$. For cosmological applications it is of particular interest to pick $R$ to be the Hubble radius. Formally, our results can be applied to any arbitrary R but, as we will discuss in more detail in section VIII, they can be physically trusted for R of the order or larger than the Hubble scale.

Then there is an alternative field operator expansion in terms of inner and outer mode functions. We restrict ourselves in the present analysis to the isotropic modes $l=0$ but it could be extended to the anisotropic modes as well.

\begin{equation}
\chi_0 = \int_0^{\infty} dk \frac{k}{\sqrt{2\omega}} \left(f_{k,in} a_{k,in} + f_{k,out} a_{k,out} + \textrm{h.c.} \right)
\end{equation}

Which means that the mode functions need to be normalized with respect to the Klein-Gordon inner product.

\begin{equation}
\begin{aligned}
& \int_0^R dr r^2 j_0(kr)j_0(kr) \sim R \\
& \int_R^L dr r^2 j_0(kr)j_0(kr) \sim L - R
\end{aligned}
\end{equation}

Where an IR regulartor $L$ has once more been introduced. We find it reasonable to suggest the following scaling for the coefficients of the mode splitting:

\begin{equation}
\alpha = \sqrt{\frac{R}{L}} \quad \textrm{\&} \quad \beta = \sqrt{1-\frac{R}{L}}
\end{equation}

As it was used in the previous section. Notice once more that $\alpha^2 + \beta^2 = 1$

The creation and annihilation operators so constructed must be treated carefully, since they do not exactly satisfy the canonical commutation relations:

\begin{equation}
\begin{aligned}
\left[ a_{k,in} , a^{\dagger}_{k',out} \right] & \sim \int_0^R drdr' rr' j_0(kr) j_0(k'r') \left[a_r,a^{\dagger}_{r'}\right]\\
& \sim \int_0^R dr r^2 j_0(kr) j_0(k'r)
\end{aligned}
\end{equation}

This integral does not give anything proportional to $\delta_{kk'}$ even though it is clerly peaked at $k=k'$. Of course, this means that the scalar product $\braket{k|k'}$ will also be proportional to this integral and, therefore, the set of states $a^n_{k,in} \ket{0}$ can be used to span the whole inner Hilbert space but it does not form an orthonormal basis. However, once the Hilbert space is restricted to one momentum mode, the set of vectors does form an orthonormal basis on that Hilbert subspace thanks to the $\delta_{nn'}$ factor appearing in the computation of the scalar product. The same applies of course to the outer Hilbert space.

These considerations do not change the form of the quantum state after inflation as we treated it in section II. The reason is that, even though a single inner or outer operator may affect several momentum modes, the combination $a_{k,in} + a_{k,out} = a_k$ does not.

One may wonder as well about the validity of the computation of the entanglement entropy, since it involves the computation of two partial traces and no orthonormal basis is available. We argue that, even though the partial traces cannot be indeed be computed exactly, the computation of section II is a good enough approximation. Let us asumme that we have at our disposal an orthonormal basis $\ket{j}$ where $j$ stands as a multi-index that labels momentum and particle number. This basis is related to our non-orthonormal basis $\ket{\tilde{j}}$ via a linear transformation:

\begin{equation}
\ket{\tilde{j}} = C \ket{j}
\end{equation}

We actually have meaningful information regarding the linear operator $C$. Its matrix elements are given by:

\begin{equation}
\begin{aligned}
C_{pqnm} & \equiv C_{jh} \equiv \bra{\tilde{j}} C \ket{h} = \braket{\tilde{j}|\tilde{h}}\\
& \sim \int_0^R dr r^2 j_0(pr)j_0(qr) \delta_{nm}
\end{aligned}
\end{equation}

The mode functions are normalized and therefore we have that $C_{jj} = 1$ and so the linear operator can be splitted into the identity plus corrections $C = 1 + \epsilon$. Since the integral is peaked at $p=q$ we assume $\epsilon$ to be small. In particular, the inverse of the operator can be written as $C^{-1} \simeq 1 - \epsilon$. Furthermore, it is traceless and so it does not affect at first order the computation of the relevant traces for our problem. Let us see how this works out for the trace of some linear operator $A$:

\begin{equation}
\begin{aligned}
\textrm{Tr} A & = \sum_{j} = \bra{j} A \ket{j} = \sum_{j'} \bra{\tilde{j}} C^{-1} A C^{-1} \ket{\tilde{j}}\\
& = \sum_{j'} \left[ \bra{\tilde{j}} A \ket{\tilde{j}} - \Re \left( \bra{\tilde{j}} A \epsilon \ket{\tilde{j}}\right) + \mathcal{O}\left(\epsilon^2\right) \right]
\end{aligned}
\end{equation}

Now let use this expression for the density matrix $\rho$ of a separable state with respect to the momentum modes such as the one created after inflation. This operator is diagonal, whereas all diagonal elements in $\epsilon$ vanish. Hence, the expected value of the product of both operators is $0$. This leaves the approximate result:

\begin{equation}
\textrm{Tr} A \simeq \sum_{\tilde{j}} \bra{\tilde{j'}} A \ket{\tilde{j}}
\end{equation}

This finishes the argument that the computation of the entanglement entropy above is a good approximation.

\section{VII. Mode counting and the area law}

The computation presented in section IV is far from accounting for the whole entanglement entropy of the region inside a sphere of radius $R$. In fact, it is limited for two reasons: it accounts only for isotropic modes ($l=0$) and only those with a given momentum $k$. Hence, it is a measure of the entanglement per isotropic mode. It is characterized by its squeezing parameter $\tau$, which is  in turn a function of the momentum $k$ and in particular the number of e-folds $N_k$ between horizon exit and the end of inflation. Roughly one gets $\tau \sim N$ \cite{Albrecht:1992kf}.

Then one simply needs to integrate:

\begin{equation}
S \sim \int dk \ \tau(k) \log R
\end{equation}

This integral could be in principle model-dependent. Notice that there is no dependence on $R^2$ as opposed to the standard area law for entanglement in QFT on 3+1 dimensions. We can understand this from the point of view that, effectively, the restriction to isotropic modes delivers a 1+1-dimensional theory. Such theories are known to have a logarithmic scaling of the entanglement entropy.

In the computation of the entanglement entropy done by Maldacena and Pimentel they also found a term proportional to the number of e-folds or, more explicitly, to $\log (-\eta)$. This computation is performed in the limit of very late time and therefore we can consider that every mode has crossed the inflationary event horizon long time ago. In that case:

\begin{equation}
\begin{aligned}
S & = \int_0^{\infty} dk N(k) \log\frac{R}{L} = \int_0^{\Lambda} dk \log(-\eta k) \log \frac{R}{L}\\
& = \Lambda \log\frac{R}{L} \left[ \log(-\eta) + \log \Lambda -1 \right]
\end{aligned}
\end{equation}

Where $\Lambda$ is a UV cut-off. In the limit $L \rightarrow \infty$ the logarithm must be replaced by a term that goes as $\sim \frac{R}{L}$ and so tends $0$. At the same time we take the limits $\Lambda \rightarrow \infty$ and keeping the product $\Lambda \frac{R}{L}$ constant. Then we get the following contributions to the entropy:

\begin{equation}
S = c \log(-\eta) + c' \log \Lambda
\end{equation}

With some coefficients $c$ and $c'$ to be determined. Both kind of terms exist in dS and therefore also in a radiation-dominated universe if we assume it is preceded by an extremely long inflationary epoch.

In order to recover the usual UV-divergent area-law scaling entanglement entropy, as well as additional UV-finite terms proportional to the area, the whole tower of $l$ and $m$ modes must be taken into account. Restricting ourselves now to the true vacuum state $\ket{0}$, it carries no angular momentum, i.e. $l=0$ and $m=0$. Angular momentum can be shown to be a good quantum number of the particle states in spherical coordinates introduced in section II. This means that $L^2 \ket{l,m} = l(l+1) \ket{l,m} $ and $L_z \ket{l,m} = m \ket{l,m}$. Therefore, if the vacuum is to be splitted, it must be done in a way that preserves the total angular momentum. This can be done with the formalism of the Clebsch-Gordan coefficients, widely used in Quantum Mechanics. One should therefore find an analogous of the singlet state of two-particle systems with spin. The difference here is that in QFT the total number of particles is not fixed \textit{a priori} and so there can be many contributions. This computation will be explored in future work.

\section{VIII. Phenomenological implications. Entangled PBH formation}

Formally, the computation showed here can be applied in principle to any entangling sphere of radius R, let it be smaller or larger than the Hubble scale $R_H$. However, from a more physical point of view, it is expected not to hold for $R<R_H$. The reason is that the modes of the scalar field and any other available quantum field defreeze after becoming sub-Hubble and start interacting. This interaction will presumably scramble the interior quantum state as well as any mode re-entering the Hubble scale at later times. This scrambling should destroy any long-range correlation inside the observable universe, although not the correlation of the observable universe with other causal domains.

The story changes if we consider some of the momentum modes to be able to trigger a gravitational collapse that creates a Primordial Black Hole \cite{GarciaBellido:1996qt, Clesse:2015wea}. The relevant scale for the formation of a Primordial Black Hole in a radiation-dominated universe is the Hubble scale, as we will briefly argue later and is supported by simple model estimates \cite{Carr:1974nx, Carr:1975qj} and numerical relativity simulations \cite{Musco:2004ak,Musco:2012au}. This means that the PBH captures most of the long-range entanglement of the Hubble sphere and keeps therefore long-range correlations with the rest of the universe, including other causal domains that collapse to form a PBH as well.

It is in this precise context that we view gravitational collapse as an \textit{entanglement trap} that prevents the long-range correlation between different Primordial Black Holes to be destroyed by scrambling. As time passes, the Hubble sphere grows and PBH formed in different causal domains come into causal contact. This creates a network of entangled PBH inside the observable universe. Note that the entanglement of super-Hubble modes arise during inflation as those modes are stretched beyond the horizon and keep this entanglement on non-causal patches. As these modes re-enter the Hubble scale after inflation and induce black hole collapse, the entanglement created during inflation is trapped inside these regions without allowing for scrambling to take place.

In other words, a PBH keeps a long-range entanglement with other PBH. This is because they trap entanglement before scrambling can take place, as scrambling is a sub-Hubble process and PBH form with a size of the order of the Hubble scale at the time of collapse. This entanglement exists regardless of whether they came into casual contact already or not. A PBH keeps a long-range entanglement as well with non-collapsed regions of the non-observable universe, as they didn’t undergo scrambling yet.

We would like to clarify that our use of entanglement entropy is not linked in principle to the gravitational entropy associated to the event horizon of any Black Hole. Instead, it is a description of how the degrees of freedom inside a spherical region are entangled with the degrees of freedom existing outside. This concept is applicable to any surface enclosing a volume. When a Black Hole is formed, the exterior degrees of freedom cannot interact with the interior ones and therefore this entanglement is preserved. It may be that the interior degrees of freedom interact with other degrees of freedom inside the Black Hole. We do not make any claim regarding the nature of the degrees of freedom inside the Black Hole, but rather than the entanglement entropy across the surface is preserved by unitarity. As an analogy, we could think of a pair of entangled photons, one of them being captured by a Black Hole and another one kept outside. It is unknown how the swallowed photon will interact with the interior degrees of freedom of the black hole, but due to unitarity the entanglement entropy of the system formed by the Black Hole and the swallowed spin will be preserved.

If this gravitational collapse is assumed to be unitary, then the entanglement entropy will be conserved during the process. Nothing forbids, for instance, the formation of a Black Hole by the collapse of a large number of particles which are entangled with distant objects. Such a Black Hole would keep this quantum entanglement. Such a process is described for instance in \cite{Maldacena:2013xja} in the context of building a pair of maximally entangled Black Holes by the gravitational collapse of the Hawking radiation of an initially isolated Black Hole.

Entangled Black Holes have been considered before in the literature \cite{Israel:1976ur, Maldacena:2013xja}, being usually maximally entangled. We have presented here a viable mechanism to produce entangled Primordial Black Holes. It must be noted, however, that they would not be maximally entangled, as their long-range entanglement entropy does not saturate the Bekenstein bound \cite{Bekenstein:1980jp}. Since two causally disconnected regions that collapse to form PBH far away from eachother are individually entangled with the rest of the universe, they must necessarily be themselves entangled with eachother. We leave for future work the computation of the fraction of long-range entanglement that actually becomes entanglement entropy between Primordial Black Holes.

It can be easily seen that $R_H$ is the relevant scale for Primordial Black  Hole formation. In a radiation-dominated universe the scale factor grows as $a \sim t^{1/2}$ and therefore the Hubble scale grows as $R_H = H^{-1} = 2t$ in natural units. With this scaling at hand, we can extract the evolution of the energy density from the second Friedmann equation:

\begin{equation}
H^2 = \frac{8\pi G}{3} \rho = \frac{1}{4t^2} \quad \textrm{and so} \quad \rho = \frac{3}{32\pi G t^2}
\end{equation}

Then it is possible to compute the mass contained inside the Hubble scale:

\begin{equation}
M = \frac{4\pi}{3} \rho (2t)^3 = \frac{t}{G}
\end{equation}

The Schwarzschild radius of a black hole of this mass corresponds precisely to the Hubble radius:

\begin{equation}
R_S = 2GM = 2t = R_H
\end{equation}

It is clear then that, up to a $\mathcal{O}(1)$ factor due to the efficiency of the gravitational collapse, the Primordial Black Hole will be of the size of the Hubble scale, i.e. $M_{\rm PBH} = \gamma M$ and so actually $R_S = \gamma R_H$. Picking $R_H$ as the radius of the entangling sphere is therefore equivalent to studying the entanglement trapped by the Primordial Black Hole.

The formation of a Primordial Black Hole by the gravitational collapse of the radiation contained inside the Hubble scale is accompanied by an enormous increase in classical entropy. Indeed, the entropy of the gas of relativistic particles within the Hubble scale can be written as \cite{Kolb:1990vq,Mukhanov:2005sc}:

\begin{equation}
S_{\rm gas} = \frac{2\pi^2}{45} g_{*S}(T) T^3 V_H
\end{equation}

Where $V_H$ is the Hubble volume, $g_*(T)$ is the number of relativistic degrees of freedom and natural units including $k_B=1$ were used, so that the entropy is a dimensionless quantity. On the other hand, the resulting Primordial Black Hole carries the Bekenstein-Hawking entropy, which is proportional to its event horizon area:

\begin{equation}
S_{\rm PBH} = \frac{A_H}{4A_P} = 4\pi \gamma^2 \frac{t^2}{t^2_P}
\end{equation}

Where $A_P = 4\pi L_P^2 $ is the Planck area, $L_P$ is the Planck length and $t_p$ is the Planck time. Since the Hubble scale is time-dependent, so are the mass and the entropy of the Primordial Black Hole.

Time and temperature are related in a radiation-dominated universe \cite{Kolb:1990vq,Mukhanov:2005sc}:

\begin{equation}
\frac{t}{t_{P}} = \left(\frac{45}{16\pi^3 g_*(T)}\right)^{1/2} \left(\frac{T_P}{T}\right)^2
\end{equation}

This way we can express both the entropy of the relativistic gas and the entropy of the Primordial Black Hole as a function of temperature:

\begin{equation}
\begin{aligned}
& S_{\rm gas} = \frac{4}{3} \frac{T_P^3}{T^3} \left(\frac{45}{16\pi^3 g_*}\right)^{1/2}\\
& S_{\rm PBH} = 4\pi \gamma^2 \left(\frac{45}{16\pi^3 g_*} \right) \frac{T_P^4}{T^4}
\end{aligned}
\end{equation}

And so the ratio of both quantities is a function of temperature as well:

\begin{equation}
\frac{S_{\rm PBH}}{S_{\rm gas}} = \left(\frac{405}{16\pi}\right)^{1/2} \gamma^2 g_*^{-1/2} \frac{T_P}{T}
\end{equation}

Let us apply this equation to the QCD phase transition temperature. Then $T\simeq 200$\,MeV and $g_* \simeq 10$. Taking into account that $T_P = 1.22 \times 10^{19}$\,GeV one gets:

\begin{equation}
\frac{S_{\rm PBH}}{S_{\rm gas}} \simeq \gamma^2 \cdot 5\times 10^{19}
\end{equation}
This large number suggests that gravitational collapse via PBH formation is an extremely efficient way of generating a burst of entropy production which could fill the universe with entropy and be alarmingly close to saturate the Bekenstein bound.

\section{IX. Conclusions and outlook}

In this paper we have studied the quantum entanglement of a scalar field during the radiation era after inflation. Thanks to the inflationary dynamics, the quantum state of the field is highly squeezed. This squeezing leads to subdominant terms in the entanglement entropy that go beyond the area-law. This kind of terms is also found in the entanglement entropy of a field living in dS and signals the survival during the radiation era of the entanglement created during inflation.

These terms arise due to the entanglement of super-Hubble modes that are stretched beyond the horizon during inflation and maintain entanglement on non-causal patches. In the case of modes that re-enter the Hubble scale after inflation and induce black hole collapse, the entanglement is trapped inside these regions without allowing for scrambling to take place.

It may seem puzzling that quantum entanglement of the state created during inflation should be conserved after its end. Indeed, if inflation is capable of creating entanglement, the next cosmological era may very likely destroy it. The creation or destruction of entanglement between quantum modes is possible since the time evolution of an individual mode can be non-unitary in presence of interactions, for instance thanks to a gravitational background. The time evolution of the total quantum state is of course unitary and remains pure. In order to gain some intuition about the survival of the entanglement, let us put in simpler, qualitative terms, the evolution that the quantum state undergoes during inflation. 

Any quantum field coupled to a gravitational background, even if minimally, is sourced by it, which leads to particle creation in the form of entangled pairs in inflation. During the radiation era, the dynamics of the field is equivalent to that of a field in Minkowski space-time and so there is no source that can affect the nature of the quantum state created during inflation.

We have assumed throughout a standard single-field inflation because of the simplicity of its treatment from a quantum field-theoretic point of view. However, we wonder whether more sophisticated models of (multi-field) inflation could enhance the entanglement. In particular, it would be fascinating if those models leading to Primordial Black Hole formation were also related to enhanced long-range entanglement. Such long-range correlations may give rise to the growth of isocurvature perturbations on cosmological scales, which could have important consequences for large scale structure formation and evolution. We intend to investigate these new phenomena in future works.

\section{Acknowledgements}

The authors acknowledge support from the Research Project FPA2015-68048-03-3P(MINECO-FEDER) and the Centro de Excelencia Severo Ochoa Program SEV-2016-0597. The work of LEP is funded by a fellowship from "la Caixa" Foundation (ID 100010434) with fellowship code LCF/BQ/IN18/11660041 and the European Union Horizon 2020 research and innovation programme under the Marie Sklodowska-Curie grant agreement No. 713673.

\bibliographystyle{h-physrev}
\bibliography{paperEPBH}

\end{document}